\documentstyle[preprint,aps,prl]{revtex}
\begin{document}
\bibliographystyle{prsty}
\draft
\title{\Large\bf Ground-state energy
and spin in disordered quantum dots}
\author{Kenji Hirose$^{1}$ and Ned S. Wingreen$^{2}$}
\address{$^{1}$ Fundamental Research Laboratories, NEC Corporation, 34
Miyukigaoka, Tsukuba, Ibaraki 305-8501, Japan\\
$^{2}$ NEC Research Institute, 4 Independence Way, Princeton, New Jersey
08540\\}
\date{\today}
\maketitle
\begin{abstract}
We investigate the ground-state energy and spin of disordered quantum 
dots using spin-density-functional theory. Fluctuations of
addition energies (Coulomb-blockade peak spacings)
do not scale with average addition energy but remain
proportional to level spacing. With increasing interaction
strength, the even-odd alternation of addition energies disappears,
and the probability of non-minimal spin increases, but never exceeds
$50\%$.
Within a two-orbital model, we show that the off-diagonal Coulomb matrix
elements help stabilize a ground state of minimal spin. 
\end{abstract}

\narrowtext
\newpage

The control of spin in semiconductor nanostructures \cite{Sohn}
is essential for a number of applications such as spintronics
\cite{Prinz} 
and quantum bits \cite{Loss}, 
and for fundamental studies of the Kondo effect in quantum dots
\cite{Kondo}.
In clean quantum dots with circular symmetry, Hund's rule is observed --
high-spin states ({\it i.e.} states of non-minimal spin)
appear for partly filled shells of degenerate 
single-particle levels \cite{Tarucha}. In disordered or chaotic quantum 
dots \cite{Alhassid}, high-spin states 
are suppressed by the 
rarity of degenerate or nearly degenerate levels. 
However, the observed
absence of even-odd alternation of addition energies in quantum dots 
\cite{Sivan,Simmel1,Patel,Simmel2} is consistent with
ground states of non-minimal spin.  
Recently, evidence for such high-spin states has been 
obtained from studies of the magnetic dispersion of Coulomb-blockade
peaks \cite{Luscher,Folk}. 
A number of theoretical treatments have considered  
the conditions under which high-spin ground states
might occur 
\cite{Prus,Blanter,Berkovits,Andreev,Brouwer,Baranger,Kurland,Jacquod}. 
Many of these works \cite{Andreev,Baranger,Kurland}
address the regime of large dimensionless conductance $g >> 1$ \cite{g};
other works rely on the validity of the Hartree-Fock approximation
\cite{Blanter,Brouwer}. 
Simulations on small lattice models indicate a significant fraction
of $S=1$ ground states \cite{Berkovits}, but we infer $g\sim 0.1-0.3$
in the simulations, compared to the experimental range $g > 1$ 
\cite{Sivan,Simmel1,Patel,Simmel2}.

Density-functional theory enables {\it in silico} experiments
on the ground-state spin of small, disordered quantum dots
for which $g > 1$, over a wide range of interaction
strengths. The results presented here
challenge the view that high-spin states 
dominate in disordered dots for $r_s > 1$ \cite{Brouwer}: 
even-odd alternation of addition energies disappears by $r_s=1.25$, 
but the probability of a high-spin ground state
never exceeds $50\%$.
Using a two-orbital model, we show that off-diagonal Coulomb
matrix elements help stabilize a ground states of minimum spin, as
argued by 
Jacquod and Stone \cite{Jacquod}. We also find that addition-energy
fluctuations do not scale with the average addition energy, but
remain proportional to the single-particle level spacing, consistent
with our previous results for spin-polarized dots \cite{Hirose2}.

The ground-state energy and spin of disordered, two-dimensional
quantum dots are obtained within spin-density-functional theory (SDFT). 
Specifically, we solve the following Kohn-Sham equations \cite{Kohn} 
numerically, and iterate until self-consistent solutions are obtained;
\begin{equation}
\left[-\frac{\hbar^2}{2m^*}\nabla^2
+\frac{e^2}{\kappa}\int\!\frac{\rho({\bf r'})}{|{\bf r-r'}|}d{\bf
r'}+\frac{\delta E_{\rm xc}[\rho,\zeta]}{\delta\rho^{\sigma}({\bf r})}
+\frac{1}{2}m^{*}\omega_0^2 r^2+V_{\rm imp}({\bf
r})\right]\Psi_i^{\sigma}
({\bf r})=\epsilon_i^{\sigma}\Psi_i^{\sigma}({\bf r}),
\label{eq:kseqs}
\end{equation}
where the density is $\rho({\bf r})=\sum_{\sigma}\rho^{\sigma}({\bf
r})=\sum_{\sigma}\sum_{i}|\Psi_i^{\sigma}({\bf r})|^2$, and the sum
is taken over the $N$ lowest-energy orbitals.
Here $\sigma$ denotes the spin index, $\zeta({\bf r})$ is the local spin
polarization, and $E_{\rm xc}[\rho,\zeta]$ is the exchange-correlation
energy functional \cite{Tanatar}. 
We use the effective mass for GaAs, $m^*=0.067m$, and 
a two-dimensional harmonic-oscillator confining potential
with $\hbar\omega_0=3.0 {\rm meV}$. The dimensionless interaction
strength is measured by
$(e^2/\kappa\ell_0)/\hbar\omega_0$ or $r_s$ \cite{rs} and is controlled
by changing the dielectric $\kappa$, where $\kappa=12.9$ for GaAs.

We use the same parameters for the total impurity potential
$V_{\rm imp}({\bf r}) =  
\sum_{i}(\gamma_{i}/2\pi\lambda^2){\rm exp}
\left(-|{\bf r}-{\bf r}_i|^2/2\lambda^2\right)$ as for a previous
study of spin-polarized dots \cite{Hirose2}. 
The density of individual impurities is $n_{\rm imp}=
1.03\times10^{-3}\, {\rm nm}^{-2}$; each has a Gaussian potential
profile, with strength $\gamma_{i}$
distributed on $[-W/2,W/2]$ with $W = 10 \hbar^2/m^*$, and width 
$\lambda=\ell_0/(2\sqrt{2})$, where $\ell_0=\sqrt{\hbar/m^*\omega_0}
\simeq 19.5 {\rm nm}$ is the quantum-confinement width of the 
parabolic potential. 
The resulting mean free path, $l=v_F\tau\simeq 120{\rm nm}$,
is comparable to the dot diameter $L = 120-160 {\rm nm}$, where
$L$ increases with interaction strength. 
Thus the dots are marginally in the ballistic
regime and have a dimensionless conductance $g \sim 2$ \cite{g}. 

The density functional method is known to give accurate ground-state 
energies and spins for a {\it clean} dot \cite{Hirose1}. Comparison with 
quantum-Monte-Carlo calculations \cite{MonteCarlo} 
confirms that SDFT is valid for
interaction strengths up to $(e^2/\kappa\ell_0)/\hbar\omega_0=6\ (r_s
\simeq 8)$ and up to 8 electrons. For disordered dots, we have 
confirmed the accuracy of the SDFT results
by exact diagonalization studies of three electrons,
over the range  $0 < (e^2/\kappa\ell_0)/\hbar\omega_0= 2.86\ (0 < r_s <
2.92)$.

At low temperatures, electron hopping into a dot is suppressed except 
when the ground-state free energies for $N-1$ and $N$ electrons are
degenerate. This condition determines the position of the $N$th
conductance peak. The $N$th peak spacing or addition energy is given by
$\Delta(N)=E(N+1)-2E(N)+E(N-1)$, where the ground-state
energy $E(N)$ is obtained from
\begin{equation}
E(N)=\sum_{i,\sigma}\epsilon_i^{\sigma}-\frac{e^2}{2\kappa}\int\frac{\rho
({\bf r})\rho({\bf r'})}{|{\bf r-r'}|}d{\bf r}d{\bf
r'}-\sum_{\sigma}\int\!
\rho^{\sigma}({\bf r})\frac{\delta E_{\rm xc}[\rho,\zeta]}
{\delta\rho^{\sigma}({\bf r})}d{\bf r}\ +\ E_{\rm xc}.
\label{eq:gsenergy}
\end{equation}

Figure 1 shows (a) the average addition energy $\langle\Delta\rangle$ 
and (b) its fluctuation
$\delta\Delta \equiv
\langle(\Delta-\langle\Delta\rangle)^2\rangle^{1/2}$,
as functions of electron-electron interaction strength
$(e^2/\kappa\ell_0)/\hbar\omega_0$ for $N=10$ and $N=11$.
Dashed curves are for total spin fixed at its minimum, $S=0\ {\rm or}\
1/2$. 
Solid curves are for unrestricted spin, {\it i.e.} the lowest 
energy $E(N)$ is found among all possible spin values.
Also shown in (a) is the average {\it noninteracting} level spacing
$\langle
\Delta_0\rangle$ and in
(b) its fluctuations $\langle \delta \Delta_0\rangle$, 
for dots of the same size as in the
interacting case. In Fig. 1(a), there is an even-odd alternation
in addition energies in the weak-interaction regime,
$(e^2/\kappa\ell_0)/\hbar\omega_0 \leq 0.95$ 
($r_s \leq 0.76$), which reflects the energy cost of adding every odd  
electron to a new orbital level. No alternation is observed
in the regime of strong interactions ($r_s > 1$) where 
$\langle\Delta\rangle$ increases nearly linearly as a function of
interaction.

In Fig. 1(b),
the magnitudes of the addition-energy fluctuations for $N=10$ and $N=11$ 
merge around $r_s \sim 1.25$. The interacting fluctuations $\delta
\Delta$
are always {\it smaller} than the fluctuations $\delta \Delta_0$ 
of the noninteracting level spacing, 
where the latter satisfy the random-matrix-theory relation 
for the Gaussian orthogonal ensemble \cite{Mehta}
$\delta\Delta_0 \simeq \sqrt{4/\pi-1}\langle\Delta_0\rangle 
\simeq 0.52\langle\Delta_0\rangle$.  The interacting fluctuations
are roughly $\delta \Delta \approx 0.28 \langle \Delta_0 \rangle$, and 
clearly do not scale with the average addition energy 
$\langle \Delta \rangle$. In fact,
the interacting fluctuations are slightly smaller than those
for the same number of interacting spin-{\it polarized}
electrons in the same potential \cite{Hirose2}.
(In the spin-polarized case, the addition-energy
fluctuations were shown to be dominated by the single-particle
level-spacing fluctuations because of the effectiveness of screening
in reducing the Coulomb fluctuations.) 

Comparing Figs. 1 (a) and (b),
one sees that in the range $r_s = 0.76 -1.25$, the average
addition energy is the same for $N=10$ and $N=11$, 
but the $N=10$ distribution has a significantly larger width. 
For $r_s = 1.25$ and below, the shapes of the $N=10$ and 
$N=11$ distributions are qualitatively different \cite{Hirose3} 
consistent with recent experiment observations for
a high-electron-density GaAs quantum dot with $r_s = 0.72$
\cite{Luscher}. 

For the strong-interaction regime $r_s > 1$, fluctuations in
addition energy are reduced when the spin states
are unrestricted.  In the lower right inset, we show
the distributions of $\Delta(N)$ for $N=10$ (upper) and for $N=11$
(lower)
for interaction strength $(e^2/\kappa\ell_0)/\hbar\omega_0=1.91
(r_s = 1.77)$, both for restricted and unrestricted spin.
The distributions of $\Delta$ are roughly Gaussian, which agrees with 
experiments at $r_s > 1$ \cite{Sivan,Simmel1,Patel,Simmel2}. (For
comparison, the Wigner-Dyson distribution\cite{Mehta} for the 
noninteracting case is shown in the upper left inset.) The 
distributions of $\Delta$ become narrower in the spin 
unrestricted case: the low-energy tail of the $N=10$ distribution is 
reduced, while the high-energy tail is reduced for $N=11$.
We trace this behavior to the appearance of $S=1$ 
ground states for 10 electrons (see below). 
Since $\Delta(N=10)=E(11)-2E(10)+E(9)$, the appearance
of new, low values for $E(10)$ removes some of the {\it lowest} values
of 
$\Delta(N=10)$. Similarly, since $\Delta(N=11)=E(12)-2E(11)+E(10)$,
new, low values of $E(10)$ remove some of the {\it highest} values 
of $\Delta(N=11)$.

Figure 2(a) shows the probabilities of the different ground-state 
spins $S$ versus
electron-electron interaction strength. Solid curves are for $N=10$
electrons (integer spin) and dashed curves are for $N=11$ electrons 
(half-integer spin). With increasing interaction strength, 
the probabilities of ground states of $S=$ 1, 3/2, and 2 increase,
while the probabilities of even higher spin states ($S=5/2$, etc.)
remain negligible. The probability of $S=1$ for $N=10$ is always higher
than that of $S=3/2$ for $N=11$. The probability of an $S=1$
ground state reaches a maximum around 
$(e^2/\kappa\ell_0)/\hbar\omega_0 \simeq 1.91 (r_s \simeq 1.77)$, {\it
and
never exceeds 50\%}. In the inset, we show the increase with interaction
strength of the probability of spin blockade, {\it i.e.} subsequent
ground
states with $|\Delta S| > 1/2$, which leads to a suppressed
Coulomb-blockade 
peak \cite{Weinmann}. 

Diagonalization studies on small lattices also find a significant
fraction of $S=1$ ground states, with a smaller likelihood of $S= 3/2$
\cite{Berkovits}. The onset of high-spin ground states
occurs at smaller $r_s \sim 0.2$, and addition-energy
fluctuations are larger, scaling as 10-20\% of the average addition
energy.
We attribute these differences to weak screening in the lattic
models due to the low dimensionless conductance
of $g = 0.1-0.3$, compared to 
$g \sim 2$ in our dots.

Berkovits \cite{Berkovits} has argued that high-spin ground states
are favored not only by exchange energy (which favors
spin alignment) but also by the enhanced Coulomb
repulsion between two electrons in the same 
spatial orbital. Opposing these
effects is the single-particle energy cost of 
promoting an electron to a new orbital. He observes that it is much
more likely to find two orbitals close in energy,
producing an $S=1$ ground state, than to find
three orbitals close in energy, as required for
an $S=3/2$ ground state.

These arguments are consistent with our SDFT results
up to $r_s \simeq 1$, but do not account for the
observed saturation of the probability of high-spin
ground states at larger $r_s$. 
To understand this saturation, 
we consider the possible states of 
two electrons occupying two spin-degenerate orbitals
near the Fermi energy. (The other electrons in the dot 
are assumed to pair up in lower-energy orbitals.)
For this two-orbital model, there are three degenerate $S=1$ states
consisting of one electron in each of the two orbitals,
and three nondegenerate $S=0$ states.

To evaluate the energies of the competing $S=0$ and $S=1$
two-electron
states, we use the basis of single-particle eigenstates
$\phi_{n}^{0}({\bf r})$ and $\phi_{n+1}^{0}({\bf r})$  
with energies $\epsilon_{n}$ and $\epsilon_{n+1}$ 
of non-interacting electrons. The various
diagonal and off-diagonal Coulomb matrix elements are
evaluated in the random-phase approximation (RPA) which
accounts for the screening effect of the other electrons in the
dot. The energy of the three degenerate $S=1$ states is
$\tilde{E}(S=1)
=\epsilon_{n}+\epsilon_{n+1}+\tilde{U}_{n,n+1}-\tilde{X}_{n,n+1}$,
where $\tilde{U}_{n,n'}=e \int
\tilde{\varphi}_{n,n}({\bf r})\rho_{n',n'}^{0}({\bf r})d{\bf r}$ is
the screened Coulomb interaction between two electrons in
orbitals $n$ and $n'$, and
$\tilde{X}_{n,n'}=e \int 
\tilde{\varphi}_{n,n'}({\bf r})\rho_{n',n}^{0}({\bf r})d{\bf r}$ is 
the screened exchange interaction.
Here, $\tilde{\varphi}_{n,n'}({\bf r})$ is the screened potential 
due to an electron, which is evaluated in Fourier representation as 
$\tilde{\varphi}_{n,n'}({\bf q})=(2\pi e/\kappa |{\bf
q}|)(\rho_{n,n'}^{0}({\bf q})/\epsilon({\bf q}))$, 
where $\rho_{n,n'}^{0}({\bf
r})=\phi_{n}^{0}({\bf r})\phi_{n'}^{0}({\bf r})$. 
The dielectric function $\epsilon({\bf q})$ is approximated as
$\epsilon^{RPA}({\bf q})=1-v_{q}\chi(q)$ where $v_{q}=2\pi/q$ 
and the susceptibility $\chi(q)=-(e^2/\kappa)(dn/d\mu)F(q/2k_F)$. The
Lindhard polarizability for 2D is $F(x)=1$ for $x \leq 1$ and
$F(x)=1-\sqrt{1-(1/x)^2}$ for $x \geq 1$.

The energy $\tilde{E}(S=0)$ of the lowest $S=0$ state is obtained by
diagonalizing the following $3 \times 3$ matrix;
\begin{equation}
\tilde{H}_{S=0}=
\left[
\begin{array}{ccc}
2\epsilon_{n}+\tilde{U}_{n,n}& \sqrt{2}\tilde{U}_{n,n,n,n+1}&
\tilde{X}_{n,n+1} \\
\sqrt{2}\tilde{U}_{n,n,n,n+1}\ \ &
\epsilon_{n}+\epsilon_{n+1}+\tilde{U}_{n,n+1}+\tilde{X}_{n,n+1}\ \ &
\sqrt{2}\tilde{U}_{n+1,n+1,n+1,n} \\
\tilde{X}_{n,n+1}& \sqrt{2}\tilde{U}_{n+1,n+1,n+1,n}&
2\epsilon_{n+1}+\tilde{U}_{n+1,n+1}
\end{array}
\right]
\label{szero}
\end{equation}
where the off-diagonal Coulomb matrix elements are 
$\tilde{U}_{n,n,n,n'}=e \int
\tilde{\varphi}_{n,n}({\bf r})\rho_{n,n'}^{0}({\bf r})d{\bf r}$.
We find that the magnitudes of the $\tilde{U}_{n,n,n,n'}$
are comparable to the exchange energy $\tilde{X}_{n,n+1}$
\cite{Jacquod}.
It is seen in Fig.~2(b) that for the two-orbital model
the average of  $\Delta\tilde{E}
\equiv \tilde{E}(S=0) - \tilde{E}(S=1)$ 
agrees reasonably well with our SDFT results
for all strengths of interaction. 
In contrast, placing the two electrons in the lowest
single-particle orbital $\phi_{n}^{0}({\bf r})$,
gives an energy $2\epsilon_{n}+\tilde{U}_{n,n}$ which is
significantly larger than $\tilde{E}(S=0)$ at 
large $r_s$.  Even the Hartree-Fock (HF)
approximation for (\ref{szero}), {\it i.e.}
the best doubly-occupied single orbital, 
overestimates the energy of the lowest 
$S=0$ state and fails to account for the saturation
of the $S=1$ probability at large $r_s$. 
It is evident that for the two-orbital model
the off-diagonal Coulomb matrix elements
help stabilize the $S=0$ ground state by creating a low-energy
hybridization of the three $S=0$ basis states,
as proposed by Jacquod and Stone \cite{Jacquod}.

We acknowledge fruitful discussions with B.~Altshuler, R.~Berkovits,
D.~Goldhaber-Gordon, B.~Halperin, C.~Marcus, M.~Stopa, and F.~Zhou.

\noindent
{\bf Figure Captions}
\begin{itemize}

\item[Figure 1:] (a) Average addition energy $\langle\Delta\rangle$ and
(b) fluctuation
$\delta\Delta=\langle(\Delta-\langle\Delta\rangle)^2\rangle^{1/2}$,
as functions of electron-electron interaction strength
$(e^2/\kappa\ell_0)/\hbar\omega_0$ for $N=10$ and $N=11$. 
Solid curves are for unrestricted spin; dashed curves are for
total spin fixed at $S=0\ {\rm or}\ 1/2$.  For each data point, 
the disorder average is taken over more than 1000 different impurity
configurations. The average noninteracting level spacing and its 
fluctuations are shown for the same size dots as in the interacting
case.
Lower right inset -- Distribution of addition energies $\Delta$ for
$N=10$ (upper) and for $N=11$ (lower) for interaction
strength $(e^2/\kappa\ell_0)/\hbar\omega_0=1.94$. 
Upper left inset -- Distribution of noninteracting
addition energies (level spacings)
for dots of the same size as in the lower inset.

\item[Figure 2:] (a) Probability of a spin $S$ ground state 
as a function of electron-electron
interaction strength $(e^2/\kappa\ell_0)/\hbar\omega_0$. 
Solid curves are for $N=10$ electrons (integer spin) and dashed 
curves are for $N=11$ electrons (half-integer spin). 
The circles give the probability of a spin $S=1$ ground state
for the full two-orbital model.
Inset -- Probability of spin blockade, {\it i.e.} subsequent 
ground states with $|\Delta S| > 1/2$.
(b) Average total energy difference $\Delta E$ between
states with $S=0$ and $S=1$ as a function of electron-electron
interaction
strength $(e^2/\kappa\ell_0)/\hbar\omega_0$. The solid curve shows
the SDFT results. Also shown are results of the two-orbital model:
exact (lowest dashed curve), Hartree-Fock (central dashed curve),
doubly-occupied lowest orbital (highest dashed curve), 
with parameters evaluated for the 5$th$ and 6$th$ non-interacting
orbitals.
The apparent agreement between the HF result and SDFT at
$r_s \sim 1$ is not meaningful -- the HF result is only
an approximation to the exact result (lowest dashed curve).
Inset -- Average level spacing and screened Coulomb interactions
for two-orbital model.

\end{itemize}


\begin{thebibliography}{1000}
\bibitem{g} The dimensionless conductance is given by
$g = \hbar/(\tau_L \langle \Delta_0\rangle)$ where the $\tau_L$ is the
time an electron takes to cross the dot and 
$\langle \Delta_0\rangle$ is the mean level spacing. For dots in the
ballistic regime, such as those studied here, 
$g \sim k_F L/\pi$ where $L$ is the linear
size of the dot.
\bibitem{rs}
The dimensionless interaction strength $r_s(=1/\sqrt{\pi\rho_0}a_B^*)$
is the
interparticle spacing in units of the effective Bohr radius $a_B^*$,
where $\rho_0$ is the electron density at the center of the dot.
\end{thebibliography}
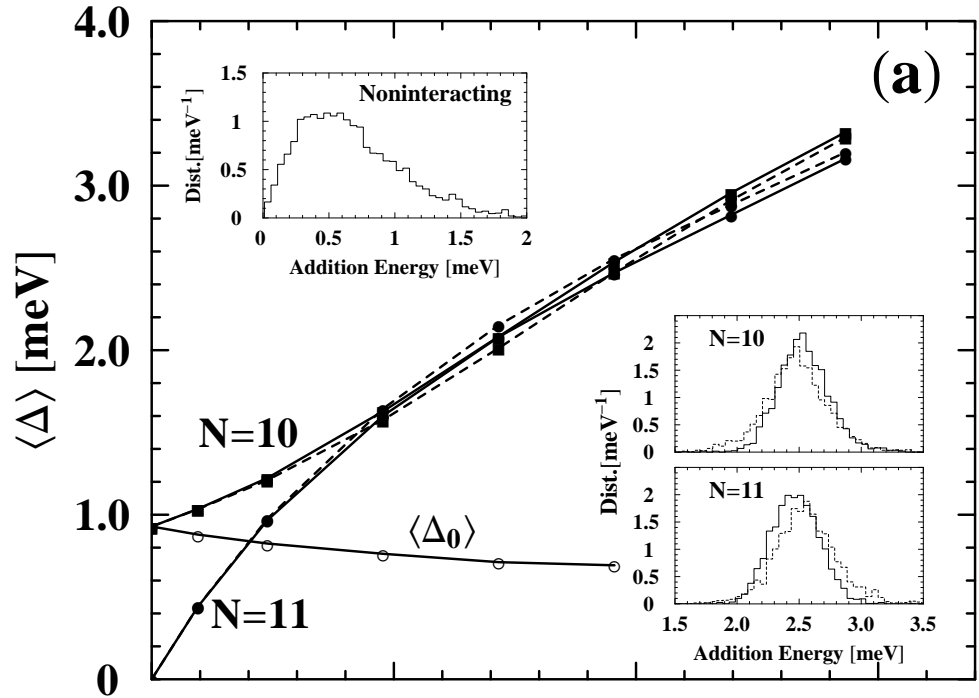
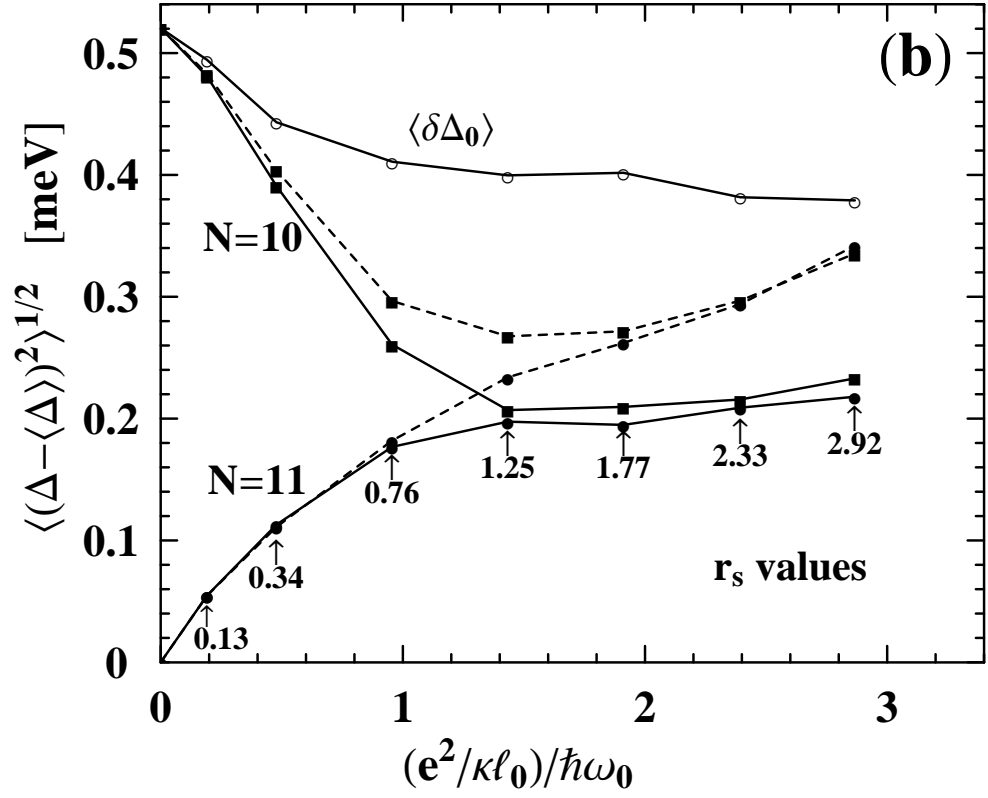
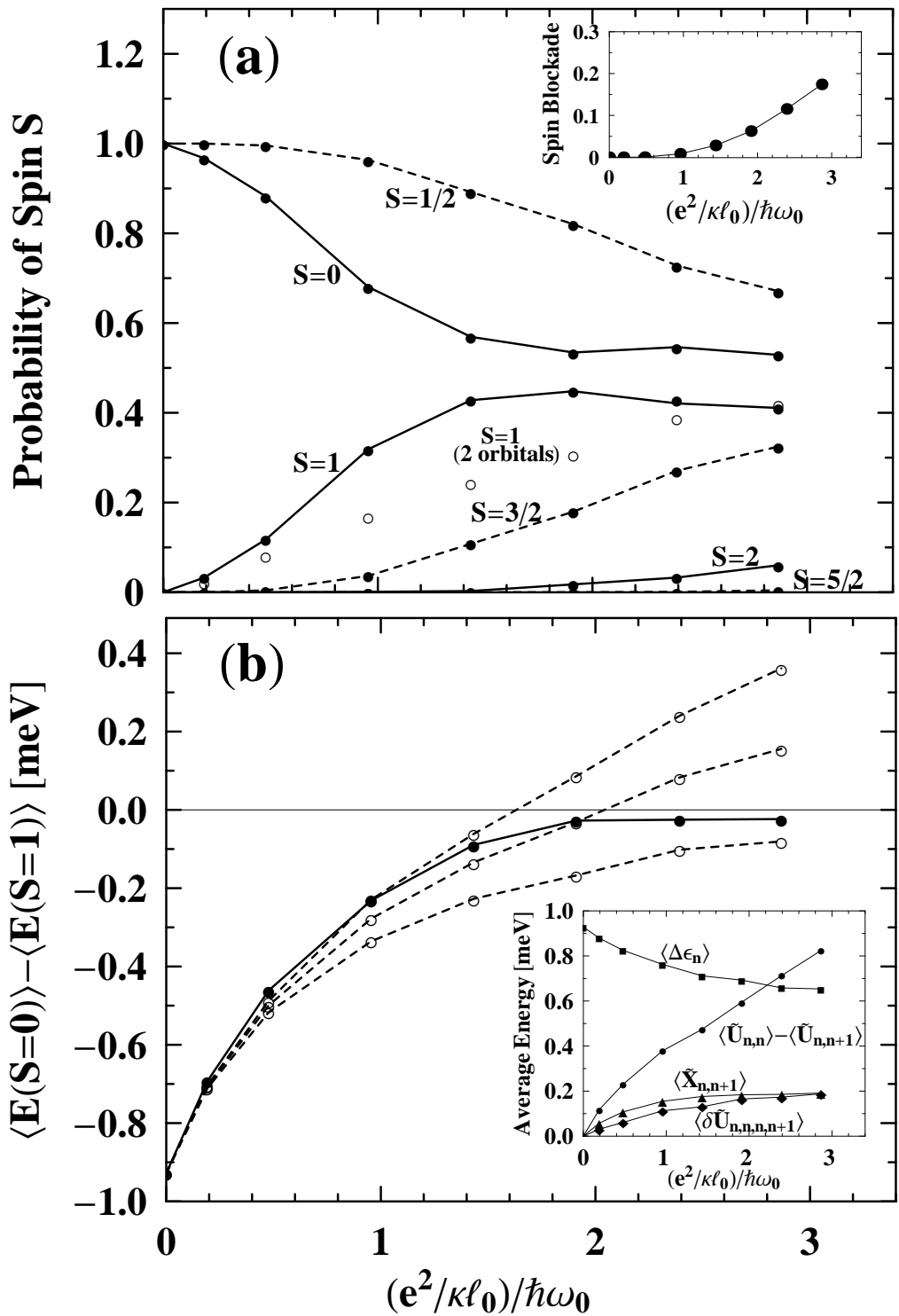
\end{document}